# Towards Parametric Speech Synthesis Using Gaussian-Markov Model of Spectral Envelope and Wavelet-Based Decomposition of F0

Mohammed Salah Al-Radhi
*Department of Telecommunications and Media Informatics*
*Budapest University of Technology and Economics*
Budapest, Hungary
malradhi@tmit.bme.hu

Tamás Gábor Csapó
*Department of Telecommunications and Media Informatics*
*Budapest University of Technology and Economics*
Budapest, Hungary
csapot@tmit.bme.hu

Csaba Zainkó
*Department of Telecommunications and Media Informatics*
*Budapest University of Technology and Economics*
Budapest, Hungary
zainko@tmit.bme.hu

Géza Németh
*Department of Telecommunications and Media Informatics*
*Budapest University of Technology and Economics*
Budapest, Hungary
nemeth@tmit.bme.hu

*Abstract*— Neural network-based Text–to–Speech has significantly improved the quality of synthesized speech. Prominent methods (e.g., Tacotron2, FastSpeech, FastPitch) usually generate Mel-spectrogram from text and then synthesize speech using vocoder (e.g., WaveNet, WaveGlow, HiFiGAN). Compared with traditional parametric approaches (e.g., STRAIGHT and WORLD), neural vocoder based end-to-end models suffer from slow inference speed, and the synthesized speech is usually not robust and lack of controllability. In this work, we propose a novel updated vocoder, which is a simple signal model to train and easy to generate waveforms. We use the Gaussian-Markov model toward robust learning of spectral envelope and wavelet-based statistical signal processing to characterize and decompose F0 features. It can retain the fine spectral envelope and achieve high controllability of natural speech. The experimental results demonstrate that our proposed vocoder achieves better naturalness of reconstructed speech than the conventional STRAIGHT vocoder, slightly better than WaveNet, and somewhat worse than the WaveRNN.

## I. INTRODUCTION

Speech synthesis is the artificial production of human speech, which is the core part of text-to-speech (TTS) systems that convert text content in a specific language into a speech waveform. It can be used in several applications, including educational, translations, telecommunications, and multimedia. Recently, statistical parametric speech synthesis (SPSS) has become a widely used speech synthesis framework due to its flexibility achieved by acoustic modeling and vocoder-based waveform generation. Hidden Markov models (HMMs) [1] and deep neural networks (DNNs) [2] have been applied to build the acoustic models for SPSS. Vocoders that reconstruct speech waveforms from acoustic features (e.g., Mel-cepstra and F0) also play an essential role in SPSS. Their performance affects the quality of synthetic speech significantly. Some conventional vocoders, such as STRAIGHT [3] and WORLD [4] designed based on the source-filter model of speech production, have been popularly applied in current SPSS systems. However, these vocoders still have some deficiencies, such as spectral details and phase information loss.

Some neural generative models for raw audio signals have been proposed and demonstrated good performance. For example, WaveNet [5] and SampleRNN [6] predicted the distribution of each waveform sample conditioned on previous samples using convolutional neural networks and recurrent neural networks (RNNs), respectively. In [7], the WaveRNN model was proposed, which generated 16-bit waveforms by splitting the RNN state into two parts and predicting the eight coarse bits and the eight fine bits. However, due to the autoregressive generation manner, these models were very inefficient at the generation stage. Therefore, some highly parallel models (e.g., parallel WaveNet [8], ClariNet [9], and WaveGlow [10]) were then proposed to improve the efficiency of generation. Experimental results confirmed that these neural vocoders performed significantly better than conventional ones. Whereas glottal neural vocoder [11], LPCNet [12], and neural source-filter (NSF) vocoder [13], have been further proposed by combining speech production mechanisms with neural networks and have also demonstrated impressive performance. However, in the current development and production scenario, it is important not only to achieve full-band and high-quality synthesis but also to allow users to control speech characteristics according to their preferences. In addition, there are still some limitations with current neural vocoders that they have much higher computation complexity than conventional STRAIGHT and WORLD vocoders. A generative adversarial network (GAN)–based excitation model has been proposed [14]. However, GANs commonly suffer from training instability and mode collapse [15]. Besides, the autoregressive neural vocoders (e.g., WaveNet, SampleRNN and WaveRNN) are very inefficient at synthesis time due to their point-by-point generation process. In contrast, the flow-based vocoders (e.g., WaveGlow) are efficient due to the flow-based model without any autoregressive connections. But, the complexity of model structures of WaveGlow is reported to be huge with low training efficiency.

To derive a simple closed-form solution, several parametrization methods exist for speech spectral modelling. These include linear predictive and cepstral coefficients, which result in a smooth spectral representation. Mel-cepstrum [16] is a well-known example of representation; it approximates the spectral envelope with a superposition of trigonometric functions. However, statistical averaging of Mel-cepstrum in SPSS changes the entire original structure and significantly degrades synthetic speech quality. To address this problem, approximation of spectral envelopes using Gaussian mixture models (GMMs) based on HMM has been proposed [17]. The GMM parameters are more stable than line spectral pair (LSP) parameters, which are other formant-related features [18].

Our recent work [19] proposed a method to decompose a multi-level representation of vocoded features using only the continuous wavelet transform. In this paper, we explore the methods to improve statistical vocoders efficiency by combining the Gaussian-Markov model toward robust learning of spectral envelope and continuous F0-based wavelet transform. The motivation of the GMM-HMM is to assign the spectral envelope of each frame to one of the hidden states that will reduce the number of the spectral features. The remarkable properties of the wavelet transform have led to powerful signal processing methods of using simple scalar transformations of individual wavelet coefficients [20]. We compare our system accuracy with the end-to-end neural models (WaveNet, WaveRNN, and NSF) and parametric vocoders (STRAIGHT and Continuous) using male and female speakers. Experimental results indicate that our framework outperforms the conventional one in quality of analysis-synthesized speech and is slightly better than the WaveNet and NSF systems. The rest of the paper is formalized as follows. First, in Section 2, the mixture of Gaussians Markov models is described, followed by a narrative of the developed fundamental frequency (F0). The experiments and simulation results are then summarized in Section 3. Finally, the conclusion is presented in Sections 4.

## II. PROPOSED METHODS

An overview of our proposed analysis-synthesis framework is shown in Figure 1.

### A. Spectral Envelope Approximation with Gaussian-Markov Model

A new analysis technique using GMM-based HMM to depict the spectral envelope of speech has been developed. We first extract spectral envelope using CheapTrick algorithm [21]. Cepstral liftering of the magnitude spectra is then applied to remove the unwanted high-quefrency effects of the excitation from the spectrum. This can yield better fits and smoother formant trajectories. After that, we approximate the modified spectral envelope with a GMM [22]. The GMM parameters are estimated by minimizing a loss function of the observed spectral envelope $H(\omega)$, and the GMM approximated one $G(\omega)$ expressed by

$$G(\omega) = \sum_{k=1}^{K} \frac{w_k}{\sqrt{2\pi\sigma_k^2}} exp\left[-\frac{(\omega - \mu_k)^2}{2\sigma_k^2}\right] \quad (1)$$

where $\omega$ denotes frequency, $K = 16$ is the number of mixture components, and $\mu_k$, $\sigma_k^2$, $w_k$ denote mean, variance, and weight of a Gaussian function with index $k$, respectively. Besides, $w_k$ is initialized with an amplitude value at the frequency of $\mu_k$ while $\sigma_k$ is initialized with a constant value. The loss function to be minimized is a divergence between two different probability distributions, and this framework uses the I-divergence $I(H, G)$ given as

$$I(H, G) = \sum_{\omega}\left[H(\omega)log\frac{Y(\omega)}{G(\omega)} - H(\omega) + G(\omega)\right] \quad (2)$$

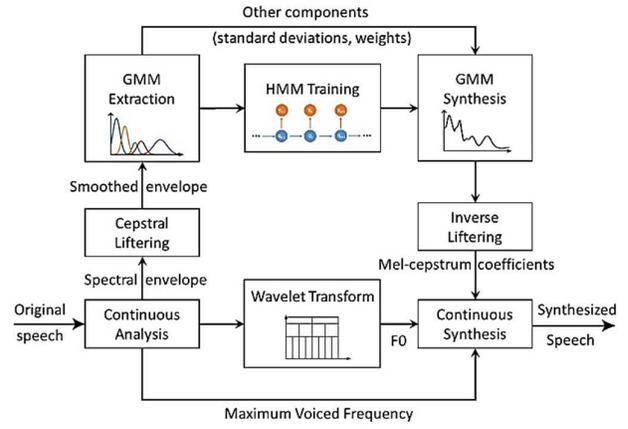

Fig. 1: Overview of proposed analysis-synthesis framework using GMM and CWT-based approximation of speech features.

Finally, we apply HMM to provide an effective framework for modelling time-varying spectral vector sequences. HMM assumes that an observation sequence was derived from a hidden state sequence of discrete data and will have associated with the means and covariances of the Gaussian distributions fit to each state. Here, HMM is fit using the Baum-Welch algorithm and decoded using the Viterbi algorithm. Let $Y_t$ be the observed variable at time $t$. The distribution of $Y_t$ depends on the state at time $t$, $X_t$. For a two-state HMM modeled by Gaussian distributions

$$f(Y_t \mid X_t = i) \sim N(\mu_i, \sigma_i^2) \quad (3)$$

$$f(Y_t \mid X_t = j) \sim N(\mu_j, \sigma_j^2) \quad (4)$$

The state at time $t$, $X_t$, depends on the previous time step, $X_t - 1$. Let $P$ be the state transition matrix, where each element $p_{i,j}$ represents the probability of transitioning from state $i$ at time $t$ to state $j$ at time $t + 1$

$$p_{ij} = P(X_t + 1 = j \mid X_t = i) \quad (5)$$

where P is a $n\ x\ n$ matrix, and $n = 2$ is the number of states of the HMM. Since we have no prior information on which to condition the first set of observations, we assume the initial probability of being in each state is the stationary distribution. We then begin with the forward step, which computes the joint probability of observing the first $t$ observations and ending up in state $i$ at time $t$, given the initial parameter estimates

$$P(X_t = i, Y_1 = y_1, Y_2 = y_2, ..., Y_t = y_t \mid \theta) \quad (6)$$

Then in the backward step, the conditional probability of observing the remaining observations after time $t$ given the state observed at time $t$ is computed

$$P(Y_t + 1 = y_t + 1, ..., Y_T = y_T \mid X_t = i, \theta) \quad (7)$$

Using Bayes' theorem, it can show that the product of the forward and backward probabilities is proportional to the probability of ending up in state $i$ at time $t$ given all of the observations

$$P(X_t = i \mid Y_1 = y_1, ..., Y_T = y_T, \theta) \quad (8)$$

## B. Wavelet-Based Decomposition of Continuous F0

Since F0 is not well defined for the unvoiced segments of the speech and the silent intervals, this makes the direct wavelet analysis impossible. Therefore, the continuous F0 is used in this study. The method in [23] significantly differs from the algorithm developed in this paper. The unvoiced gaps in [23] were filled using traditional linear interpolation, which is different from our method. We applied a continuous pitch estimation algorithm [24] which used: 1) Bayesian approach that naturally yields estimates for unvoiced segments, along with variances for all estimates; 2) Kalman smoother to the sequence of estimates and variances to give a sequence of pitch estimates.

A wavelet is a short waveform with finite duration, whose average value is zero. The continuous wavelet transform (CWT) can describe the signal in various transformations of a mother wavelet. Scaling the mother wavelet, the transform can capture high frequencies if the wavelet is compressed and low frequencies if it is stretched. The process is repeated by translating the mother wavelet. The CWT output is an $MxN$ matrix where $M$ is the number of scales and $N$ is the length of the signal. The CWT coefficient at scale $a$ and position $b$ is given by:

$$W(a,b) = \frac{1}{\sqrt{a}} \int_{-\infty}^{+\infty} f(x)\psi(\frac{x-b}{a})dx \qquad (9)$$

where $x$ is the input signal, and $\psi$ is the mother wavelet. A set of 10 components is defined, where each component is approximately one octave apart. The original signal can be recovered from the wavelet representation by inverse transform using the double-integral form over all scales and locations, $a$ and $b$ (for the proof, see [25])

$$f(x) = \int_{-\infty}^{+\infty}\int_{-\infty}^{+\infty} \frac{W(a,b)}{a^2\sqrt{a}}\psi(\frac{b-x}{a})dxda \qquad (10)$$

Then we can obtain an approximation to the original signal by summing the scaled CWT coefficients over all scales

$$f(x) = \sum_{i=1}^{10} W_i(a,b;x) + \epsilon(x) \qquad (11)$$

where $\epsilon(x)$ is the reconstruction error. CWT can analyze a speech waveform with a time-frequency resolution different from the Short-time Fourier Transform. This leads to high-frequency resolution with CWT at low frequencies and high time resolution at high frequencies. In Figure 2, the second pane shows the CWT of the F0 contour of an English sentence shown in the bottom pane.

## III. EXPERIMENTS

### A. Experimental Setup

To evaluate of the proposed system, we used two speakers from the CMU-ARCTIC database [26], where SLT is female and BDL is male. The sampling frequency is set to 16 kHz with 16-bit linear quantization. The total number of utterances is 1132 per speaker, and the entire utterance duration is about 1 hour per speaker. Acoustic features were extracted every 5ms after applying a window of 25ms. In the experiment, we compare our proposed model with the following speech methods:

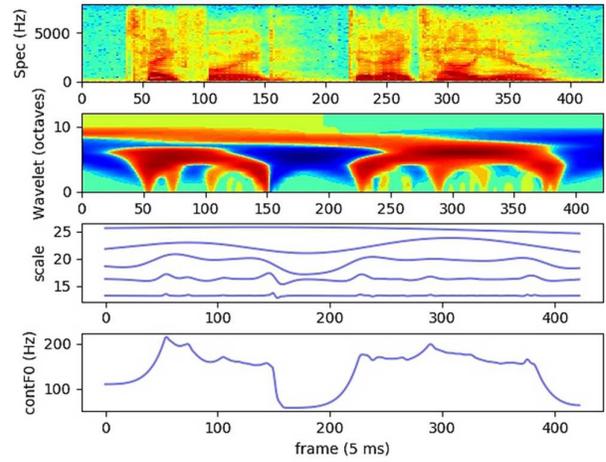

Fig. 2: Top pane shows a spectrogram of the speech signal, the second pane depicts the continuous wavelet transform with Mexican hat mother wavelet of F0, the third pane shows the scales, and bottom pane gives the modified continuous F0.

- **WaveNet:** It was trained by using 80-dim log-Mel spectrograms. The network architecture of the WaveNet was the same as that used in [8]. The total number of utterances is 6580 for training and 350 for testing (about 6 hours of recorded speech).

- **NSF:** The model structure and the training method of the NSF vocoder were the same as that of the phase spectrum predictor model in the HiNet vocoder [13].

- **WaveRNN:** The structure of a 16-bit WaveRNN-based neural vocoder implemented here was the same one used in [7]. The built model had one hidden layer of 1024 nodes where 512 nodes for coarse outputs and another 512 nodes for acceptable results. The waveform samples were quantized to discrete values by 16-bit linear quantization.

- **STRAIGHT:** The conventional STRAIGHT vocoder. At synthesis time, the spectral envelope at each frame was first reconstructed from input Mel-cepstra and frame energy, and was then used to generate speech waveforms together with input source parameters [3].

- **Continuous (Baseline):** It was used as a baseline vocoder [27] in this work. The total number of dimensions of the continuous vocoder were 26 (24 Mel-cepstrum + 1 MVF + 1 contF0).

- **Anchor:** For lower anchor, we used a simple pulse-noise excitation vocoder. Before the synthesis, the 24-order spectral components [28] were distorted (the 13-24th features were replaced with their average), thus resulting in a low quality resynthesis.

### B. Evaluation Metrics

To prove the proposed model is correct and accurate, we evaluated Mel-Cepstrum Distortion (MCD) between the natural speech and synthesized speech:

$$MCD = \frac{10}{\log 10}\sqrt{\sum_{m=1}^{M}\left(c_{org}(m) - c_{synth}(m)\right)^2} \quad (12)$$

where $c_{org}$ and $c_{synth}$ are Mel-cepstrum from original and synthesized speech, respectively, and $M$ is the order of Mel-cepstrum. The average MCD results from the natural and synthesized speech are presented in Table 1.

TABLE 1. COMPARISON OF MEL-CEPSTRUM DISTORTION BETWEEN SPECTRAL FEATURES OF NATURAL AND SYNTHESIZED SPEECH.

| Systems | MCD (dB) | |
|---|---|---|
| | *Male* | *Female* |
| Baseline | 4.086 | 4.194 |
| STRAIGHT | 3.792 | 3.925 |
| NSF | 3.671 | 3.650 |
| WaveNet | 3.785 | 3.924 |
| WaveRNN | 3.428 | 3.589 |
| Proposed | 3.399 | 3.564 |

Comparing the proposed method with the baseline and other end-to-end systems, the suggested model decreases the value of MCD, proving that the GMM-HMM has a significant impact on the spectral feature of speech synthesis. This means that it could reproduce the original spectrum correctly. Thus, the proposed method could capture the spectral information with relatively higher accuracy and outperform the state-of-the-art neural vocoders.

*C. Subjective Results*

To calculate the perceptual quality of the developed method, we performed a web-based MUSHRA (MUlti-Stimulus test with Hidden Reference and Anchor) listening experiment. We evaluated natural sentences with the synthesized ones from the baseline, STRAIGHT, NSF, WaveNet, WaveRNN, proposed, and an anchor system. The participants had to assess the naturalness of each stimulus relative to the reference (which was the natural sentence), from 0 (highly unnatural) to 100 (highly natural). Ten participants (7 males, 3 females) with a mean age of 33 years and no known hearing defects were invited to run the online perceptual test. On average, the MUSHRA test took 10 minutes. The listening test samples are available online http://smartlab.tmit.bme.hu/eusipco2022

Results are presented in Figure 3. As can be seen, there are still differences when compared against the WaveRNN system. However, the proposed method is significantly preferred over the baseline and achieves higher naturalness than the WaveNet and NSF neural vocoders (Mann-Whitney-Wilcoxon ranksum test with a 95% confidence level). Hence, our method presents a good alternative technique to other systems for the reconstruction of speech.

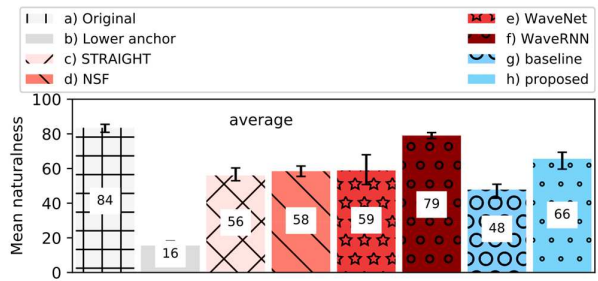

Fig. 3: Sound quality of synthesized speech.

IV. CONCLUSIONS

The current paper has proposed a speech analysis-synthesis system based on the Gaussian-Markov model of spectral envelope and wavelet-based decomposition of F0. Objective metrics and sound quality tests supported our work. We confirmed through experiments that our speech model could generate a natural-sounding synthetic speech and superior to state-of-the-art WaveNet vocoder on the CMU-ARCTIC database. This means that our new system is simple and requires fewer acoustic parameters than the neural vocoder investigates in our paper. Future work includes optimizing the number of components in the DNN-based TTS and voice conversion frameworks.

ACKNOWLEDGMENT

The research reported in this publication, carried out by Department of Telecommunications and Media Informatics Budapest University of Technology and Economic and IdomSoft Ltd., was supported by the Ministry of Innovation and Technology and the National Research, Development and Innovation Office within the framework of the National Laboratory of Infocommunication and Information Technology. The research was partly supported by the APH-ALARM project (contract 2019-2.1.2-NEMZ-2020-00012) funded by the European Commission and the National Research, Development and Innovation Office of Hungary. Tamás Gábor Csapó's research was supported by the Bolyai János Research Fellowship of the Hungarian Academy of Sciences and by the ÚNKP-21-5 (identifier: ÚNKP-21-5-BME-352) New National Excellence Program of the Ministry for Innovation and Technology from the source of the National, Research, Development and Innovation Fund.